\newcommand{\bea}{\begin{eqnarray}}
\newcommand{\eea}{\end{eqnarray}}
\newcommand{\bear}{\begin{eqnarray*}}
\newcommand{\eear}{\end{eqnarray*}}

\documentstyle[aps,preprint]{revtex}
\begin{document}

\draft

\title
{PHASE DIAGRAM AND SUPERCONDUCTING PROPERTIES OF AN  EXACTLY SOLVABLE MODEL
WITH CORRELATED HOPPING }

\author{
F. C. Alcaraz$^1$ and R. Z. Bariev$^{1,2}$}

\address{$^1$Departamento de F\'{\i}sica, 
Universidade Federal de S\~ao Carlos, 13565-905, S\~ao Carlos, SP
Brazil}

\address{$^2$The Kazan Physico-Technical Institute of the Russian Academy of Sciences, 
Kazan 420029, Russia}

\maketitle

\begin{abstract}
The model of strongly correlated electrons with the correlated hopping term  
and an additional interaction between holes $V$ is solved exactly 
in one dimension at a special point  where the number of hole pairs is 
conserved. 
As a function of the interaction $V$ the phase diagram has a rich structure. 
There exist gapless phases with dominating density-density correlations 
(metallic) or pair-pair correlations (superconducting) as well insulating 
 and phase-separated phases.
 The stiffness constant of the conductivity, the effective transport mass, 
and the compressibility are also exactly calculated in several regions of 
the phase diagram.
\end{abstract}


\narrowtext
The study of strongly correlated fermion systems has attracted considerable
attention during the last decade, in particular due to the discovery of
high-temperature superconductivity. Exactly solvable models, mainly those 
obtained using the Bethe-ansatz, have brought a very important progress in 
this area. Two well-studied models, which are exactly integrable  
in one dimension, are the Hubbard
model and $t-J$ model [1-5]. 
In [6] a new model was proposed, 
which
is again solvable in one dimension, and it was shown [7] to be  
relevant to the theory of "hole"
superconductivity introduced by Hirsh [8]. According to this theory  
the introduction of a correlated hopping term in the 
Hubbard Hamiltonian leads to an attractive effective
interaction between the holes and the formation of Cooper pairs. Although
this picture was confirmed by a BCS-type mean-field theory it is desirable 
to find exact results confirming this behavior. A first step in this direction
has been made in [7] where a simplified version of Hirsch's Hamiltonian 
 has been
studied in one dimension.  It was  found that this model has a strong
tendency towards superconductivity although it does not have finite  
off-diagonal long-range order.

Another interesting general problem in the physics of highly correlated 
electronic systems is the characterization of the metallic, insulating and 
superconducting phases [9,10]. In this sense it is desirable to introduce exactly 
integrable models showing these phases.  In this letter we introduce a  
model where all these phases are present. This model along with a 
correlated hopping term $t_1$ also has a static interaction $V$ between 
holes.
We show 
that this model is integrable provided that $t_1 = t$,  where $t$ is the 
standard hopping matrix element, and study the
superconducting properties at this point analytically. The phase diagram 
is very rich with a crossover from a region
exhibiting  long-range  order, which corresponds to phase separation, 
to a regime with 
a complicated antiferromagnetic order, that corresponds to an insulating phase. 
Gapless  regions with dominating
superconducting pair correlation functions or 
dominating density-density correlations without long-range order also exist.

The Hamiltonian of the correlated hopping model with an additional interaction
beetween holes on a one-dimensional lattice with $L$ sites and periodic
boundary conditions is given by

\bea
H(t_1,V) = &-& \sum_{j=1}^L\sum_{\sigma=\pm 1}
\lbrack \left(c_{j,\sigma}^+c_{j+1,\sigma} 
+c_{j+1,\sigma}^+ c_{j,\sigma}
 \right)\nonumber\\
&\times &\left(1 - t_1 n_{j+(1+\sigma)/2,-\sigma}\right)\nonumber\\
 &+& V\left(2-n_{j,\sigma}-n_{j+1,\sigma}
 \right)\left(1 - n_{j,-\sigma}\right)\rbrack,
\eea
where $c_{j,\sigma}$ is the annihilation operator for an electron with spin
$\sigma$ at site $j$ and $n_{j,\sigma} = c_{j,\sigma}^+c_{j,\sigma}$. Here we
choose  the hopping matrix element $t = 1$. The integrable 
correlated hopping chain [7] is a special case of (1) with $V = 0$.

Firstly we will be interested in the special case $H(V) = H(t_1 = 1, V)$. Note
that the analogous construction have been considered for the Hubbard model
in [10-12]. The results of this consideration are quite different from ours.

In the Hamiltonian $H(V)$ only one type of processes is allowed : a hole at 
the site $(j,\sigma)$ may hop to an occupied neighbour site $(j\pm 1, \sigma)$
provided that the site $(j + (\sigma \pm 1)/2, -\sigma)$ is empty. 
This process corresponds to the motion of the hole pair on the half 
of the lattice
distance. In the Hamiltonian $H(V)$ there is also the static interaction 
between holes which corresponds to an extremely anisotropic  spin-spin 
interaction [12]. We consider hereafter the situation where we have only pair 
of holes [13].
 In this 
case we consider hole pairs as "particles" and treat this problem within 
the framework of the Bethe-ansatz method.

The discrete Bethe-ansatz equations are derived following the standard 
procedure by imposing periodic boundary conditions. Each state of the 
Hamiltonian is specifed by the set of charge rapidities $p_j$ representing the 
momenta of the hole pairs. All rapidities within a given set have to be 
different corresponding to Fermi statistics. These rapidities are determined by the Bethe-ansatz equations

\bea
e^{2iLp_j} =(-1)^{m-1}\prod_{l=1}^m 
\exp \left[i(p_j - p_l)-i\Theta (p_j,p_l) \right] ;\nonumber\\
\exp \left[-i\Theta (p_j,p_l) \right] = S(p_j,p_l)/S(p_l,p_j),\nonumber \\
S(p_j,p_l)=\left(1-2Ve^{ip_j} + e^{ip_j + ip_l}\right).
\eea
The energy and  momentum of the corresponding state are given by 

\bea
E = -2\sum_{j =1}^m\left(\cos p_j + V-\mu\right)-2L\mu,\hspace{0.5cm}
P = \sum_{j =1}^m p_j;
\eea
where the chemical potential $\mu$ has been added to control the particle
 number.

The above equations are valid  for density of holes $\rho = m/L \leq 2/3$. 
For $\rho > 2/3 $ we can choose as the 
reference state the state with unoccupied 
sites and treat again this problem by the Bethe-ansatz method. Each state
of the Hamiltonian is now specified by the set of charge 
rapidities $k_j$ representing the momenta of the electrons. 
These rapidities are determined by the following Bethe-ansatz equations

\bea
e^{iLk_j} = - \prod_{l=1}^n\exp\lbrack\frac{i}{2}(k_l-k_j)-
i\Theta (k_j,k_l)\rbrack .
\eea
The corresponding energy and momentum are given by 

\bea
E = -4LV + \sum_{j=1}^n \left(4V -\mu - 2\cos k_j\right), \hspace{0.5cm} P =
 \sum_{j=1}^n k_j.
\eea
We may consider (2,3) and (4,5) in the straighforward method [14]. 
In the
thermodynamic limit (2) and (4) are replaced by the integral equations for the 
distribution functions
\bea  
2\pi R(U) = K_1(U)
\left(1 -\frac{\rho}{2}\right)
 - \int_{-U_0}^{U_0} K_2(U-U')R(U')dU', \nonumber 
\eea 
\bea
\int_{-U_0}^{U_0}R(U)dU = \cases{
\frac{1}{2}\rho,& $\rho \leq 2/3$,\cr
\left(1 - \rho \right),& $2/3 \leq \rho \leq 1$,\cr}
\eea
where for $ -1 < V = -\cos\gamma < 1 $
\bea
K_{\alpha}(U) = \sin (\alpha\gamma)[\cosh U - \cos(\alpha\gamma)]^{-1}, 
(\alpha = 1,2 ) \nonumber
\eea
and for $ V = -\cosh\lambda < -1$
\bea
K_{\alpha}(U) = \sinh (\alpha\lambda)[\cosh U - \cos(\alpha\lambda)]^{-1}, 
(\alpha = 1,2 ). \nonumber 
\eea
The energy is given by 
\bea
\frac{1}{2L} E = \int_{-U_0}^{U_0}\epsilon(U)dU
\eea
where  the dressed energy $\epsilon(U)$ is the solution of the 
integral equation
\bea
\epsilon(U) = 
\epsilon_0(U) - \frac{1}{2\pi}\int_{-U_0}^{U_0}K_2(U-U')\epsilon(U')dU'.
\eea
Here for $V < 1$
\bea 
\epsilon_0(U) = -2\sin\gamma K_1(U) + \frac{1}{2}(1\pm 3)
(\mu + 2\cos\gamma), \nonumber
\eea
with sign $+$ ($-$) for  $\rho < 2/3$  ($\rho > 2/3$), respectively. For  
$V < -1$
\bea
\epsilon_0(U) = -2\sinh\lambda K_1(U) + 2(\mu + 2\cosh\lambda).
\nonumber
\eea
In all cases $\epsilon(\pm U_0) = 0$.

At $V < -1$ and $\rho > 2/3$, there is a gap in the energy spectrum and it corresponds to the existence of long-range order in the system.
 In this case the equations
(6,8) are not valid.

At $V>1$ there is a gap for the arbitrary concentration of holes and this 
state has long-range order which corresponds to phase separation at 
$V = -1$.

For $-1 < V < 1$ there is no gap in the eigenspectrum 
and in order to understand the superconducting properties of the model 
we shall investigate 
the long-distance behavior of the correlation functions. For this 
purpose we  shall use  two-dimensional conformal field theory [15] and 
 analytic methods [16,17] 
to extract finite-size corrections from the Bethe-ansatz 
equations.
The results of these calculations are the following.

The long-distance power-law behavior of the density-density
correlation functions is given by

\bea
\left<\rho(r)\rho(0)\right>\simeq\rho^2+A_1r^{-2}+A_2r^{-\alpha}
\cos(2k_Fr),   
\eea
where
\bea
 2k_F=\pi\rho, \hspace{0.5cm}
\rho(r)&=&\sum_{\sigma}c_{r\sigma}^+c_{r\sigma}. \nonumber
\eea
The critical exponent $\alpha$ is 
expressed in terms of the dressed charge $\xi(U_0)$ which  satisfies 
the integral equation 
\bea
\xi(U) + \frac{1}{2\pi}\int_{-U_0}^{U_0}
K_2(U-U') \xi(U')dU' = 
1 - \frac{1}{2}\rho,
\eea
The Eq. (10) and (14) add to the integral equations (6) and (8) [18].  
The superconducting properties of the system 
manifest themselves in the behavior of the pair 
correlation function 
\bea
G_p(r)&=&\left<c_{r\uparrow}^+c_{r,\downarrow}^+c_{0\downarrow}c_{0\uparrow}
\right>\simeq Br^{-\beta} , \nonumber\\
\beta& =& \alpha^{-1} = 
\frac{1}{2[\xi(U_0)]^2}. 
\eea

The exponent $\beta$ is plotted in Fig.1 for some values of the parameter $V$.  
These results are obtained by solving  numerically the integral equations 
(6) and (10). We see from this figure that for all values of $V$ there exists 
a critical density $\rho_c$  separating a regime with dominant density-density 
correlations $\beta > 1$ from a regime with dominant pair superconducting 
 correlations 
$\beta < 1$. These two regimes are represented in the phase diagram of Fig. 2 
by the phases $C$ and $B$, respectively. The critical curve $\rho_c(V)$ is 
obtained numerically and is represented by the broken line. In phase $B$, 
where the superconducting correlation function decays slower than 
the density-density correlation function we have the closest analogy with 
a true superconducting phase. Analogous behavior of correlation functions 
are observed in other models [7,19], but differently from these models, 
in the present case we have the nice feature that for increasing values of 
$\rho$, the exponent $\beta$ tends toward zero. This means that the system 
try to have a true off-diagonal long-range-superconducting order (ODLRSO). 
However 
at the same time when $\beta \rightarrow 0$, which happens for $V \rightarrow 
1^-$, the hole compressibility, as we shall see, tends to infinity and phase 
separation takes place. Then at $V = 1$ we have two phase transitions 
happening simultaneously. We may expect that in two dimensions these phase 
transitions happen at different places [20].

In region $D$ ($V < -1, \rho >2/3$) we have no bound pairs but  a long-range 
"antiferromagnetic order", where  hole pairs and electrons are located 
alternately. This is an insulating gapped phase. In order to see this we 
calculated the electrical conductivity in our model. This can be done by 
studying the Hamiltonian (1) with twisted boundary condition, where the 
twisting angle $\phi$ corresponds to an enclosed magnetic flux in the 
ring [9]. The conductivity is directly proportional to the charge stifness 
$D_c$, which can be obtained from the change $\Delta E_0 = D_c\phi^2/L$ of 
the ground-state energy $E_0$, for small $\phi$ and large $L$. From the charge 
stiffness we can define an effective transport mass by the relation
\bea
m/m_e = D_c^0/D_c ,
\eea
where $D_c^0 = 2\sin(\pi \rho)/\pi$ is the charge stiffness of the 
non-interacting system and $m_e$ is the electron mass. Extending the Bethe 
ansatz equations (2-5) for the case of twisted boundary conditions [21] 
$c_{L+1,\sigma}^{\pm} = \exp (\pm i \phi_{\sigma})c_{1,\sigma}^{\pm}$ with 
$\phi_+ =\phi_- = \phi$ and exploring the finite size corrections obtained 
from conformal invariance [15] we obtain, after some calculation
\bea
D_c = 2v_F \xi^2/\pi, \; v_F = \epsilon'(U_0)/(4\pi R(U_0)),
\eea
where by prime we mean derivative. In Fig. 3 we show $m/m_e$ obtained 
numerically for several values of $V$. We clearly see in this figure that 
for $V < -1$ as $\rho \rightarrow 2/3$ the transport mass diverges, 
which indicates that phase $D$ in Fig. 2 is an insulating phase, and at 
$\rho = 2/3$ we have a Mott transition. 
The anomalous behavior in Fig. 1 and 3 around $\rho = 2/3$ at $V = -0.98$ 
is due to the strong competition between the repulsive and kinetic energy, 
which for larger values of $V$ will lead to the Mott transition. 
This figure also shows that for 
$-1 < V < 1$ the carrier mass is around $2m_e$ for moderate densities, which 
means that the conductivity is ruled by the motion of pairs.

For $V > 1$ (phase $A$ in Fig. 2) we have bound pairs, for any density and 
there is a long-range order in the system. This is a region where phase 
separation takes place. The hole pairs are separated from the particles. 
In order to see this we calculated the hole compressibility $\kappa = 
1/(\rho^2 \frac{d^2E_{\rho}}{d \rho^2})$, where $E_{\rho}$ is the 
ground-state energy for a given concentration $\rho$. Exploring, as before, 
the finite-size corrections obtained from conformal invariance [15] we obtain 
$\kappa = D_c/2v_F^2 \rho^2$. In Fig. 4 we show the  curves of $\kappa$ for 
certain values of $V$. We clearly see in this figure that $\kappa$ increases 
drastically for all densities as we tend toward $V =1$, indicating 
a phase separation for $V > 1$ (phase $A$ in Fig. 2) [22].

Thus we have investigated a correlated hopping model [7] with an additional 
hole-hole interaction $V$ at the special point $t_1 = 1$ where  the
number of hole pairs is conserved and we have only hole pairs. 
We have shown that our model exhibits phase separation and Mott transition 
at $V = 1$ and $V = -1$, respectively. 
At $-1 < V <1$ there are regions with dominating superconducting
 correlations. Although the system in this region has no ODLRSO it is 
very close to the true superconducting state since the critical exponent of 
the superconducting correlation function $\beta$ tends to zero.

In the general case at $0 < t_1 < 1$ desintegration of hole
pairs is possible, and the number of hole pairs is not conserved. 
However these states
are not favorable in energy and their contribution is 
negligible  since they are separated from those considered above 
by a large energy gap, 
 at least for a small density of holes. As a result, we may expect that 
at arbitrary $ t_1 $ the qualitative behavior of the system will be same as 
that considered  at 
$ t_1 = 1 $.  Certainly it is desirable to investigate this problem 
using perturbation theory and exact diagonalization of small systems.

This work was supported in part by Conselho Nacional de Desenvolvimento 
Cient\'{\i}fico e Tecnol\'ogico  - CNPq - Brazil, 
and by Funda\c c\~ao de Amparo \`a Pesquisa 
do Estado de S\~ao Paulo - FAPESP - Brazil.

\begin{figure}[tbp]
\caption{  
 The exponent $\beta$ of the pair correlation function as a function 
of hole density $\rho$ for some values of the interaction $V$.}
\end{figure}

\begin{figure}[tbp]
\caption{  
 Phase diagram as a function of the interaction $v$ and hole 
density $\rho$. Phases $A$ and $D$ are gapped and correspond to phase 
separations and insulating phases, respectively. Phases $B$ and $C$ are gapless 
and corresponds to the region with dominating behavior of superconducting 
correlations and density-density correlations, respectively.}
\end{figure}

\begin{figure}[tbp]
\caption{  
 The effective transport mass as a function of density of hole $\rho$ 
for some values of $V$.}
\end{figure}

\begin{figure}[tbp]
\caption{  
The hole compressibility  as a function of density of hole $\rho$ 
for some values of $V$.}
\end{figure}

\end{document}